\documentclass{article}
\usepackage{ijcai19}

\usepackage{booktabs} 
\usepackage{graphicx}
\usepackage{balance}  
\usepackage{amssymb}
\usepackage{graphicx}
\usepackage{float}
\usepackage{subfigure}
\usepackage{mathtools}
\usepackage{eurosym}

\usepackage{pgfplots}
\usepackage{url}
\usepackage{enumitem}
\usepackage[linesnumbered,ruled]{algorithm2e}
\usepackage[export]{adjustbox}
\usepackage{xspace}
\usepackage{breqn}
\usepackage{flushend}
\newcommand{\separate}{\vspace{6pt}}
\newcommand{\separateshort}{\vspace{3pt}}

\title{Exploration of Interesting Dense Regions in Spatial Data}

\author{
Pl\'acido A. Souza Neto$^1$\and
Francisco B. Silva J\'unior$^1$\and \\
Felipe F. Pontes$^1$\and
Behrooz Omidvar-Tehrani$^2$\\
\affiliations
$^1$Federal Institute of Rio Grande do Norte (Brazil)\\
$^2$University of Grenoble Alpes (France)\\
\emails
placido.neto@ifrn.edu.br,
bento.francisco@academico.ifrn.edu.br,
freire.pontes@academico.ifrn.edu.br,
behrooz.omidvar-tehrani@univ-grenoble-alpes.fr
}

\begin{document}

\maketitle

\begin{abstract}
Nowadays, spatial data are ubiquitous in various fields of science, such as transportation and the social Web. A recent research direction in analyzing spatial data is to provide means for ``exploratory analysis'' of such data where analysts are guided towards interesting options in consecutive analysis iterations. Typically, the guidance component learns analyst's preferences using her explicit feedback, e.g., picking a spatial point or selecting a region of interest. However, it is often the case that analysts forget or don't feel necessary to explicitly express their feedback in what they find interesting. Our approach captures implicit feedback on spatial data. The approach consists of observing mouse moves (as a means of analyst's interaction) and also the explicit analyst's interaction with data points in order to discover interesting spatial regions with dense mouse hovers. In this paper, we define, formalize and explore Interesting Dense Regions (IDRs) which capture preferences of analysts, in order to automatically find interesting spatial highlights. Our approach involves a polygon-based abstraction layer for capturing preferences. Using these IDRs, we highlight points to guide analysts in the analysis process. We discuss the efficiency and effectiveness of our approach through realistic examples and experiments on \textsf{Airbnb} and \textsf{Yelp} datasets.
\end{abstract}

\section{Introduction}
Nowadays, there has been a meteoric rise in the generation of spatial datasets in various fields of science, such as transportation, lodging services, and social science. As each record in spatial data represents an activity in a precise geographical location, analyzing such data enables discoveries grounded on facts. Analysts are often interested to observe spatial patterns and trends to improve their decision making process. Spatial data analysis has various applications such as smart city management, disaster management and autonomous transport~\cite{RoddickEHPS04,Telang:2012}.

\separateshort
Typically, spatial data analysis begins with an imprecise question in the mind of the analyst, i.e., {\em exploratory analysis}. The analyst requires to go through several trial-and-error iterations to improve her understanding of the spatial data and gain insights. Each iteration involves visualizing a subset of data on geographical maps using an  off-the-shelf product (e.g., Tableau\footnote{\it http://www.tableau.com}, Exhibit\footnote{\it http://www.simile-widgets.org/exhibit/}, Spotfire\footnote{\it http://spotfire.tibco.com}) where the analyst can investigate on different parts of the visualization by zooming in/out and panning.

\separateshort
Spatial data are often voluminous. Hence the focus in the literature of spatial data analysis is on ``efficiency'', i.e., enabling fluid means of navigation in spatial data to facilitate the exploratory analysis. The common approach is to design pre-computed indexes which enable efficient retrieval of spatial data (e.g., \cite{lins2013nanocubes,yu2018spatial}). However, there has been less attention to the ``value'' derived from spatial data. Despite the huge progress on the efficiency front, an analyst may easily get lost in the plethora of geographical points due to two following reasons.

\separate
\noindent $\blacksquare$ In an exploratory context, the analyst doesn't know a priori what to investigate next.

\separateshort
\noindent $\blacksquare$ Moreover, she may easily get distracted and miss interesting points by visual clutter caused by huge point overlaps.

\separate
The main drawback of the traditional analysis model is that the analyst has a {\em passive role} in the process. In other words, the analyst's feedback (i.e., her likes and dislikes) is ignored and only the input query (i.e., her explicit request) is served. In case feedback is incorporated, the process can be more directed towards analyst's interests where her partial needs can be served earlier in the process. In this paper, we advocate for a ``guidance layer'' on top of the raw visualization of spatial data to enable analysts know {\em ``what to see next''}. This guidance should be a function of analyst feedback: the system should return options similar to what the analyst has already appreciated. 

\separateshort
Various approaches in the literature propose methodologies to incorporate analyst's feedback in the exploration process of spatial data~\cite{Ballatore2008,Liu:2010,xin2006discovering,bhuiyan2012interactive}. Typically, feedback is considered as a function which is triggered by any analyst's action on the map. The action can be ``selecting a point'', ``moving to a region'', ``asking for more details'', etc. The function then updates a ``profile vector'' which keeps tracks of analyst's interests. The updated content in the profile vector enables the guidance functionality. For instance, if the analyst shows interest in a point which describes a house with balcony, this choice of amenity will reflect her profile to prioritize other houses with balcony in future iterations.

\separateshort
Feedback is often expressed {\em explicitly}, i.e., the analyst clicks on a point and mentions if she likes or dislikes the point \cite{kamat2014distributed,Omidvar-Tehrani:2015,omidvar2017geoguide}. In \cite{omidvar2017geoguide}, we proposed an interactive approach to exploit such feedback for enabling a more insightful exploration of spatial data. However, there are several cases that the feedback is expressed {\em implicitly}, i.e., the analyst does not explicitly click on a point, but there exist correlations with other signals captured from the analyst which provide hint on her interest. For instance, it is often the case in spatial data analysis that analysts look at some regions of interest but do not provide an explicit feedback. Another example is frequent mouse moves around a region which is a good indicator of the analyst's potential interest in the points in that region. Implicit feedbacks are more challenging to capture and hence less investigated in the literature. The following example describes a use case of implicit feedbacks. This will be our running example which we follow throughout the paper.

\separateshort
\noindent {\bf Example.} {\em Ben\'icio is planning to live in Paris for a season. He decides to rent a home-stay from \textsf{Airbnb} website\footnote{\it http://www.airbnb.com}. He likes to discover the city, hence he is open to any type of lodging in any region with an interest to stay in the center of Paris. The website returns 1500 different locations. As he has no other preferences, an exhaustive investigation needs scanning each location independently which is nearly infeasible. While he is scanning few first options, he shows interest in the region of Trocadero (where the Eiffel tower is located at) but he forgets or doesn't feel necessary to click a point there. An ideal system should capture this implicit feedback in order to short-list a small subset of locations that Ben\'icio should consider as high priority}.

\separateshort
The above example shows in practice that implicit feedback capturing is crucial in the context of spatial data analysis. While text-boxes, combo-boxes and other input elements are available in analyzing other types of data, the only interaction means between the analyst and a spatial data analysis system is a geographical map spanned on the whole screen. In this context, a point can be easily remained out of sight and missed.

\separateshort
In this paper, we present an approach whose aim is to capture and analyze implicit feedback of analysts in spatial data analysis. Without loss of generality, we focus on ``mouse moves'' as the implicit feedback received from the analyst. Mouse moves are the most common way that analysts interact with geographical maps~\cite{Chen:2001}. It is shown in \cite{Arapakis:2014} that mouse gestures have a strong correlation with ``user engagement''. Intuitively, a point gets a higher weight in the analyst's profile if the mouse cursor moves around it frequently.  However, our approach can be easily extended to other types of inputs such as gaze tracking, leap motions, etc.

\separate
\noindent {\bf Contributions.} In this paper, we make the following contributions:

\begin{itemize}[leftmargin=*]
  \item We define and explore the notion of ``implicit user feedback'' which enables a seamless navigation in spatial data;
  \item We define the notion of ``information highlighting'', a mechanism to highlight out-of-sight important information for analysts. A clear distinction of our proposal with the literature is that it doesn't aim for pruning (such as top-k recommendation), but leveraging the actual data with potential interesting results (i.e., highlights);
  \item  We define and formalize the concept of Interesting Dense Regions (IDRs), a polygon-based approach to explore and highlight spatial data;
  \item We propose an efficient greedy approach to compute highlights on-the-fly;
  \item We show the effectiveness of our approach through a set of qualitative experiments.
\end{itemize}  

\separateshort
The outline of the paper is the following. Section \ref{sec:datamodel} describes our data model. In Section \ref{sec:problem}, we formally define our problem. Then in Section \ref{sec:algo}, we present our solution and its algorithmic details. Section  \ref{sec:experiments} reports our experiments on the framework. We review the related work in Section \ref{sec:rel}. We present some limitations of our work in Section \ref{sec:limitations}. Last, we conclude in Section~\ref{sec:conclusion}.

\section{Data Model}
\label{sec:datamodel}
We consider two different layers on a geographical map: ``spatial layer'' and ``interaction layer''. The spatial layer contains points from a spatial database $\mathcal{P}$. The interaction layer contains mouse move points $\mathcal{M}$.

\separate
\noindent {\bf Spatial layer.} Each point $p \in \mathcal{P}$ is described using its coordinates, {\em latitude} and {\em longitude}, i.e., $p = \langle \mathit{lat}, \mathit{lon} \rangle$. Note that in this work, we don't consider ``time'' for spatial points, as our contribution focuses on their location. Points are also associated to a set of domain-specific attributes $\mathcal{A}$. For instance, for a dataset of a real estate agency, points are properties (houses and apartments) and $\mathcal{A}$ contains attributes such as ``surface'', ``number of pieces'' and  ``price''. The set of all possible values for an attribute $a \in \mathcal{A}$ is denoted as $dom(a)$. We also define analyst's feedback $F$ as a vector over all attribute values (i.e., facets), i.e., $F = \overrightarrow{\cup_{a \in \mathcal{A}}dom(a)}$. The vector $F$ is initialized by zeros and will be manipulated to express analyst's preferences.

\separate
\noindent {\bf Interaction layer.} Whenever the analyst moves her mouse, a new point $m$ is appended to the set $\mathcal{M}$. Each mouse move point is described using the pixel position that it touches and the clock time of the move. Hence each mouse move point is a tuple $m = \langle x, y, t \rangle$, where $x$ and $y$ specifies the pixel location and $t$ is a Unix Epoch time. To conform with geographical standards, we assume $m = \langle 0, 0\rangle$ sits at the middle of the interaction layer, both horizontally and vertically.

\separateshort
The analyst is in contact with the interaction layer. To update the feedback vector $F$, we need to translate pixel locations in the interaction layer to latitudes and longitudes in the spatial layer. While there is no precise transformation from planar to spherical coordinates, we employ equirectangular projection to obtain the best possible approximation. Equation \ref{eq:equirectangular} describes this formula to transform a point $m = \langle x,y,t \rangle$ in the interaction layer to a point $p = \langle lat, lon \rangle$ in the spatial layer. Note that the resulting $p$ is not necessarily a member of $\mathcal{P}$. 

\begin{equation}\label{eq:equirectangular}
\mathit{lon} = \frac{x}{\mathit{cos}\gamma} + \theta; \mathit{lat} = y + \gamma 
\end{equation}

The inverse operation, i.e., transforming from the spatial layer to the interaction is done using Equation \ref{eq:reverse}.

\begin{equation}\label{eq:reverse}
x = (\mathit{lon} - \theta) \times \mathit{cos}\gamma; y = \mathit{lat} - \gamma
\end{equation}

\separateshort
The reference point for the transformation is the center of both layers. In Equations \ref{eq:equirectangular} and \ref{eq:reverse}, we assume that $\gamma$ is the latitude and $\theta$ is the longitude of a point in the spatial layer corresponding to the center of the interaction layer, i.e., $m= \langle 0,0 \rangle$.

\section{Problem Definition}
\label{sec:problem}
The large size of spatial data hinders its effective analysis for discovering insights. Analysts require to obtain only few options (so-called ``highlights'') to focus on. These options should be in-line with what they have already appreciated. In this paper, we formulate the problem of ``information highlighting using implicit feedback'', i.e., highlight few spatial points based on implicit interests of the analyst in order to guide her towards what she should concentrate on in consecutive iterations of the analysis process. We formally define our problem as follows.

\separate
\noindent {\bf Problem.} {\em Given a time $t_c$ and an integer constant $k$, obtain an updated feedback vector $F$ using points $m \in \mathcal{M}$ where $m.t \leq t_c$ and choose $k$ points $\mathcal{P}_k \subseteq \mathcal{P}$ as ``highlights'' where $\mathcal{P}_k$ satisfies two following constraints.}

\separateshort
\noindent $\blacksquare$ $\forall p \in \mathcal{P}_k, \mathit{similarity}(p,F)$ {\em is maximized.}

\separateshort
\noindent $\blacksquare$ $\mathit{diversity}(\mathcal{P}_k)$ {\em is maximized.}

The first constraint guarantees that returned highlights are highly similar with analyst's interests captured in $F$. The second constraint ensures that $k$ points cover different regions and they don't repeat themselves. While our approach is independent from the way that {\em similarity} and {\em diversity} functions are formulated, we provide a formal definition of these functions in Section \ref{sec:algo}.

\separateshort
The aforementioned problem is hard to solve due to the following challenges.

\separate
\noindent $\blacksquare$ {\bf Challenge 1.} First, it is not clear how mouse move points influence the feedback vector. Mouse moves occur on a separate layer and there should be some meaningful transformations to interpret mouse moves as potential changes in the feedback vector. 

\separateshort
\noindent $\blacksquare$ {\bf Challenge 2.} Even if an oracle provides a mapping between mouse moves and the feedback vector, analyzing all generated mouse moves is challenging and may introduce false positives. A typical mouse with 1600 DPI (Dots Per Inch), touches 630 pixels for one centimeter of move. Hence a mouse move from the bottom to the top of a typical 13-inch screen would provide 14,427 points which may not be necessarily meaningful.

\separateshort
\noindent $\blacksquare$ {\bf Challenge 3.} Beyond two first challenges, finding the most similar and diverse points with $F$ needs an exhaustive scan of all points in $\mathcal{P}$ which is prohibitively expensive: in most spatial datasets, there exist millions of points. Moreover, we need to follow multi-objective considerations as we aim to optimize both similarity and diversity at the same time.

\separate
We recognize the complexity of our problem using the aforementioned challenges. In Section \ref{sec:algo}, we discuss a solution for the discussed problem and its associated challenges.

\section{Interesting Dense Regions}
\label{sec:algo}
Our approach exploits analyst's implicit feedback (i.e., mouse moves) to highlight few interesting points as future analysis directions. Algorithm \ref{algo:main} summarizes the principled steps of our approach.

\begin{algorithm}[t]
\DontPrintSemicolon
\KwIn{Current time $t_c$, mouse move points $\mathcal{M}$}
\KwOut{Highlights $\mathcal{P}_k$}
$\mathcal{S} \gets \mathit{find\_interesting\_dense\_regions}(t_c,\mathcal{M})$\label{ln:dense}\;
$\mathcal{P}_s \gets \mathit{match\_points}(\mathcal{S}, \mathcal{P})$\label{ln:match}\;
$F \gets \mathit{update\_feedback\_vector}(F, \mathcal{P}_s)$\label{ln:update}\;
$\mathcal{P}_k \gets \mathit{get\_highlights}(\mathcal{P}, F)$\label{ln:highlight}\;
\Return{$\mathcal{P}_k$}\; 
\caption{Spatial Highlighting Algorithm}
\label{algo:main}
\end{algorithm}

\separateshort
The algorithm begins by mining the set of mouse move points~$\mathcal{M}$ in the interaction layer to discover one or several Interesting Dense Regions, abbr., IDRs, in which most analyst's interactions occur (line \ref{ln:dense}). Then it matches the spatial points $\mathcal{P}$ with IDRs using Equation \ref{eq:reverse} in order to find points inside each region (line~\ref{ln:match}). The attributes of resulting points will be exploited to update the analyst's feedback vector~$F$~(line \ref{ln:update}). The updated vector $F$ will then be used to find $k$ highlights (line \ref{ln:highlight}). These steps ensure that the final highlights reflect analyst's implicit interests. We detail each step as follows.

\subsection{Discovering IDRs}
The objective of this step is to obtain one or several regions in which the analyst has expressed her implicit feedback. There are two observations for such regions.

\separate
\noindent $\blacksquare$ {\bf Observation 1.} We believe that a region appeals more interesting to the analyst if it is denser, i.e., the analyst moves her mouse in that region several times.

\separateshort
\noindent $\blacksquare$ {\bf Observation 2.} It is possible that the analyst moves her mouse everywhere in the map. This should not signify that everywhere in the map has the same significance.

\separate
Following our observations, we propose Algorithm \ref{algo:dense} for mining IDRs. We add points to $\mathcal{M}$ only every $200ms$ to prevent adding redundant points (i.e., Challenge 2).  Following Observation 1 and in order to mine the recurring behavior of the analyst, the algorithm begins by partitioning the set $\mathcal{M}$ into $g$ fixed-length consecutive segments $\mathcal{M}_0$ to $\mathcal{M}_g$. The first segment starts at time zero (where the system started), and the last segment ends at $t_c$, i.e., the current time. Following Observation 2, we then find dense clusters in each segment of $\mathcal{M}$ using a variant of DB-SCAN approach~\cite{Ester:1996}. Finally, we return intersections among those clusters as IDRs.

\separateshort
For clustering points in each time segment (i.e., line \ref{ln:mine} of Algorithm~\ref{algo:dense}), we use ST-DBSCAN~\cite{Birant:2007}, a space-aware variant of DB-SCAN for clustering points based on density. For each subset of mouse move points $\mathcal{M}_i$, $i \in [0,g]$, ST-DBSCAN begins with a random point $m_0 \in \mathcal{M}_i$ and collects all density-reachable points from $m_0$ using a distance metric. As mouse move points are in the 2-dimensional pixel space (i.e., the display), we choose euclidean distance as the distance metric. If $m_0$ turns out to be a core object, a cluster will be generated. Otherwise, if $m_0$ is a border object, no point is density-reachable from $m_0$ and the algorithm picks another random point in $\mathcal{M}_i$. The process is repeated until all of the points have been processed.

\separateshort
Once clusters are obtained for all subsets of $\mathcal{M}$, we find their intersections to locate recurring regions (line \ref{ln:poly}). To obtain intersections, we need to clearly define the spatial boundaries of each cluster. Hence for each cluster, we discover its corresponding polygon that covers the points inside. For this aim, we employ Quickhull algorithm, a quicksort-style method which computes the convex hull for a given set of points in a 2D plane~\cite{Barber:1996}.

\begin{figure*}[t]
\centering
   \includegraphics[width=\textwidth]{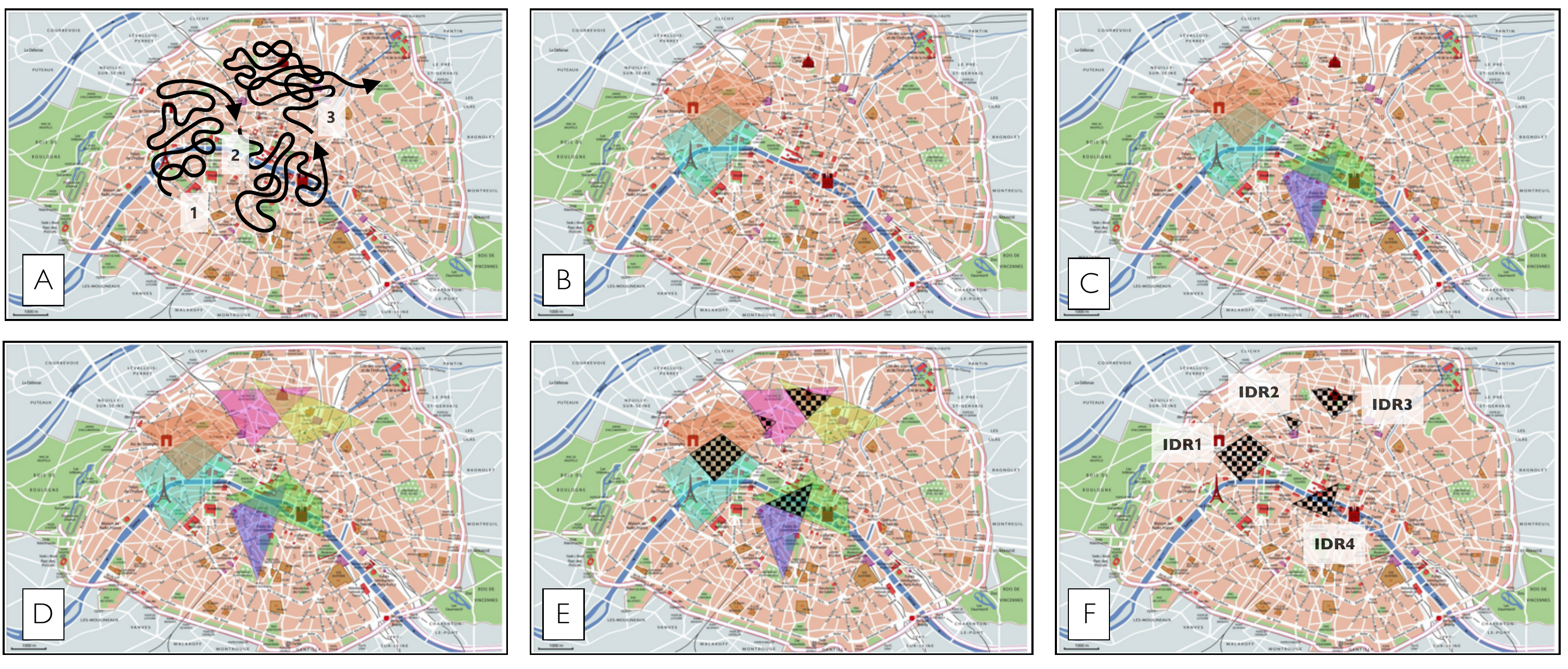}
  \caption{The process of finding IDRs on \textsf{Airbnb} dataset.}
  \label{fig:regions}
\end{figure*}

\begin{algorithm}[t]
\DontPrintSemicolon
\KwIn{Current time $t_c$, mouse move points $\mathcal{M}$}
\KwOut{IDRs $\mathcal{S}$}
$\mathcal{S} \gets \emptyset$\;
$g \gets ${\em number of time segments}\;
\For{$i \in [0,g]$}
{
       $\mathcal{M}_i \gets \{m = \langle x,y,t \rangle | (\frac{t_c}{g} \times i) \leq t \leq (\frac{t_c}{g} \times (i+1))\}$\;
       $\mathcal{C}_i \gets \mathit{mine\_clusters}(\mathcal{M}_i)$\label{ln:mine}\;
       $\mathcal{O}_i \gets \mathit{find\_ploygons}(\mathcal{C}_i)$\label{ln:poly}\;
}
\lFor{$\mathcal{O}_i, \mathcal{O}_j$ where $i,j \in [0,g]$ and $i \neq j$}
{
       $\mathcal{S}.\mathit{append}(\mathit{intersect}(\mathcal{O}_i, \mathcal{O}_j))$
}
\Return{$\mathcal{S}$}\; 
\caption{Find Interesting Dense Regions (IDRs)}
\label{algo:dense}
\end{algorithm}

\begin{figure}[t]
\centering
   \includegraphics[width=\columnwidth]{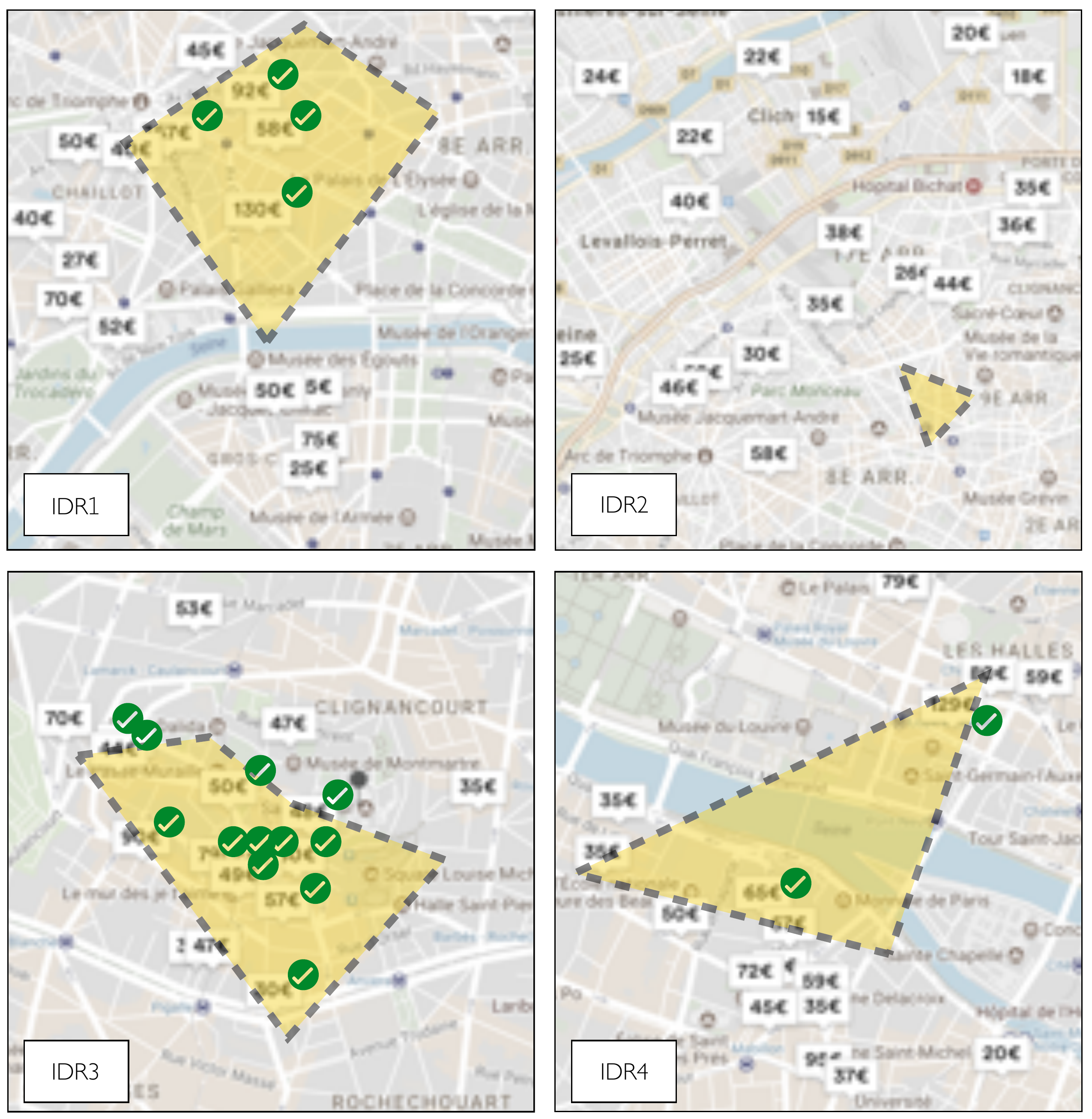}
  \caption{Matching points for IDR1 to IDR4.}
  \label{fig:match}
\end{figure}

\separateshort
We describe the process of finding IDRs in an example. Figure~\ref{fig:regions} shows the steps that Ben\'icio follows in our running example to explore home-stays in Paris. Figure \ref{fig:regions}.A shows mouse movements of Ben\'icio in different time stages. In this example, we consider $g = 3$ and capture Ben\'icio's feedback in three different time segments (progressing from Figures \ref{fig:regions}.B to \ref{fig:regions}.D). It shows that Ben\'icio started his search around Eiffel Tower and Arc de Triomphe (Figure \ref{fig:regions}.B) and gradually showed interest in south (Figure \ref{fig:regions}.C) and north (Figure \ref{fig:regions}.D) as well. All intersections between those clusters are discovered~(hatching regions in Figure \ref{fig:regions}.E) which will constitute the set of IDRs (Figure \ref{fig:regions}.F), i.e., IDR1 to IDR4.

\subsection{Matching Points}
Being a function of mouse move points, IDRs are discovered in the interaction layer. We then need to find out which points in $\mathcal{P}$ fall into IDRs, hence forming the subset $\mathcal{P}_s$. We employ Equation \ref{eq:reverse} to transform those points from the spatial layer to the interaction layer. Then a simple ``spatial containment'' function can verify which points fit into the IDRs. Given a point $p$ and an IDR $r$, a function $\mathit{contains}(p,r)$ returns ``true'' if $p$ is inside $r$, otherwise ``false''. In our case, we simply use the implementation of $\mathit{ST\_Within}(p,r)$ module in PostGIS\footnote{\it https://postgis.net/docs/manual-dev/ST\_Within.html}, i.e., our underlying spatial DBMS which hosts the data.

\separate
In the vanilla version of our spatial containment function, all points should be checked against all IDRs. Obviously, this depletes the execution time. To prevent the exhaustive scan, we employ Quadtrees~\cite{finkel1974quad} in a two-step approach.

\separateshort
\noindent $\blacksquare$ In an offline process, we build a Quadtree index for all points in $\mathcal{P}$. We record the membership relations of points and cells in the index.

\separateshort
\noindent $\blacksquare$ When IDRs are discovered, we record which cells in the Quadtree index intersect with IDRs. As we often end up with few IDRs, the intersection verification performs fast. Then for matching points, we only check a subset which is inside the cells associated to IDRs and ignore the points outside. This leads to a drastic pruning of points in $\mathcal{P}$.

\separate
We follow our running example and illustrate the matching process in Figure \ref{fig:match}. In the \textsf{Airbnb} dataset, points are home-stays which are shown with their nightly price on the map. We observe that there exist many matching points with IDR3 and absolutely no matching point for IDR2. For IDR4, although there exist many home-stays below the region, we never check their containment, as they belong to a Quadtree cell which doesn't intersect with the IDR. 

\subsection{Updating Analyst Feedback Vector}
The set of matching points $\mathcal{P}_s$ (line \ref{ln:match} of Algorithm \ref{algo:main}) depicts the implicit preference of the analyst. We keep track of this preference in a feedback vector $F$. The vector is initialized by zero, i.e., the analyst has no preference at the beginning. We update $F$ using the attributes of the points in $\mathcal{P}_s$.

\separateshort
We consider an {\em increment value} $\delta$ to update $F$. If $p \in \mathcal{P}_s$ gets $v_1$ for attribute $a_1$, we augment the value in the $F$'s cell of $\langle a_1, v_1 \rangle$ by $\delta$. Note that we only consider incremental feedback, i.e., we never decrease a value in~$F$.

\separateshort
We explain the process of updating the feedback vector using a toy example. Given the four matched points in IDR1 (Figure \ref{fig:match}) with prices 130\euro, 58\euro, 92\euro\ and 67\euro, we want to update the vector~$F$ given those points. Few attributes of these points are mentioned in Table \ref{tbl:attribs}. In practice, there are often more than 50 attributes for points. The cells of $F$ are illustrated in the first column of Table \ref{tbl:feedback}. As three points get the value ``1'' for the attribute ``\#Beds'', then the value in cell $\langle$\#Beds,1$\rangle$ is augmented three times by $\delta$. The same process is repeated for all attribute-values of points in $\mathcal{P}_s$. Note that all cells of $F$ are not necessarily touched in the feedback update process. For instance, in the above example, 5 cells out of 12 remain unchanged.

\separateshort
By specifying an increment value, we can materialize the updates and normalize the vector using a Softmax function. We always normalize $F$ in a way that all cell values sum up to $1.0$. Given $\delta = 1.0$, the normalized values of the $F$ vector is illustrated in the third column of Table \ref{tbl:feedback}. Higher values of $\delta$ increase the influence of feedbacks.

\separateshort
The normalized content of the vector $F$ captures the implicit preferences of the analyst. For instance, the content of $F$ after applying points in IDR1 shows that the analyst has a high interest in having a balcony in her home-stay, as her score for the cell $\langle$Balcony,Yes$\rangle$ is 0.25, i.e., the highest among other cells. This reflects the reality as all points in IDR1 has balcony. Note that although we only consider positive feedback, the Softmax function lowers the values of untouched cells once other cells get rewarded.

\separateshort
An important consideration in interpreting the vector $F$ is that the value ``0'' does not mean the lowest preference, but {\em irrelevance}. For instance, consider the cell $\langle$Rating,2$\rangle$ in Table \ref{tbl:feedback}. The value ``0'' for this cell shows that the analyst has never expressed her implicit feedback on this facet. It is possible that in future iterations, the analyst shows interest in a 2-star home-stay (potentially thanks to its price), hence this cell gets a value greater than zero. However, cells with lower preferences are identifiable with non-zero values tending to zero. For instance, the value 0.06 for the cell $\langle$Rating,4$\rangle$ shows a lower preference towards 4-star home-stays compared to the ones with 5 stars, as only one point in $\mathcal{P}_s$ is rated 4 in IDR1.

\begin{table}[t]
\centering
\caption{Attributes of points in IDR1.}
\label{tbl:attribs}
\begin{tabular}{|c|c|c|c|c|c|}
\hline
\textbf{ID} & \textbf{Price} & \textbf{\#Bed} & \textbf{Balcony} & \textbf{Air-con.} & \textbf{Rating} \\ \hline
1                     & 130\euro           & 1               & Yes           & Yes                & 5/5             \\ \hline
2                     & 58\euro            & 1               & Yes           & No                 & 5/5             \\ \hline
3                     & 92\euro            & 2               & Yes           & No                 & 5/5             \\ \hline
4                     & 67\euro            & 1               & Yes           & No                 & 4/5             \\ \hline
\end{tabular}
\end{table}

\begin{table}[t]
\centering
\caption{Updating Analyst Feedback Vector}
\label{tbl:feedback}
\begin{tabular}{|c|c|c|}
\hline
\textbf{Attribute-value}               & \textbf{Applying IDR 1} & \textbf{Normalized} \\ \hline
$\langle$\#Beds,1$\rangle$                   & $+3\delta$                       & 0.19                 \\ \hline
$\langle$\#Beds,2$\rangle$                 & $+\delta$                       & 0.06                 \\ \hline
$\langle$\#Beds,+2$\rangle$                  & {\em (no update)}                       & 0.00                    \\ \hline
$\langle$Balcony,Yes$\rangle$                   & $+4\delta$                      & {\bf 0.25}                 \\ \hline
$\langle$Balcony,No$\rangle$                    & {\em (no update)}                        & 0.00                    \\ \hline
$\langle$Air-cond.,Yes$\rangle$               & $+\delta$                       & 0.06                 \\ \hline
$\langle$Air-cond.,No$\rangle$                & $+3\delta$                       & 0.19                 \\ \hline
$\langle$Rating,1$\rangle$                    & {\em (no update)}                       & 0.00                    \\ \hline
$\langle$Rating,2$\rangle$                     & {\em (no update)}                        & 0.00                    \\ \hline
$\langle$Rating,3$\rangle$                    & {\em (no update)}                        & 0.00                   \\ \hline
$\langle$Rating,4$\rangle$                   & $+\delta$                       & {\bf 0.06}                 \\ \hline
$\langle$Rating,5$\rangle$                     & $+3\delta$                      & 0.19                 \\ \hline
\end{tabular}
\end{table}

\subsection{Generating Highlights}
The ultimate goal is to highlight $k$ points to guide analysts in analyzing their spatial data. The updated feedback vector $F$ is the input to the highlighting phase. We assume that points in IDRs are already investigated by the analyst. Hence our search space for highlighting is $\mathcal{P} - \mathcal{P}_s$.

\separateshort
We seek two properties in $k$ highlights: {\em similarity} and {\em diversity}. First, highlights should be in the same direction of the analyst's implicit feedback, hence similar to the vector~$F$. The similarity between a point $p \in \mathcal{P}$ and the vector~$F$ is defined as follows.

\begin{equation}
       \label{eq:rel}
       \mathit{similarity}(p,F) = \mathit{avg}_{a \in \mathcal{A}}(\mathit{sim(p, F, a)})
\end{equation}

\separateshort
The $\mathit{sim}()$ function can be any function such as Jaccard or Cosine. Each attribute can have its own similarity function (as string and integer attributes are compared differently.) Then $\mathit{sim}()$ works as an overriding-function which provides encapsulated similarity computations for any type of attribute.

\separateshort
Second, highlighted points should also represent distinct directions so that the analyst can observe different aspects of data and decide based on the big picture. Given a set of points $\mathcal{P}_k =  \{ p_1, p_2 \dots p_k \}  \subseteq{\mathcal P} $, we define { \em diversity} as follows.

\begin{equation}
       \label{eq:divs}
       \mathit{diversity}(\mathcal{P}_k) = \mathit{avg}_{\{p, p'\} \subset \mathcal{P}_k | p \neq p' } \mathit{distance}(p,p')
\end{equation} 

\separateshort
The function $\mathit{distance}(p,p')$ operates on geographical coordinates of $p$ and $p'$ and can be considered as any distance function of Minkowski distance family. However, as distance computations are done in the spherical space, a natural choice is to employ Haversine distance shown in Equation~\ref{eq:harvestine}. Our application of diversity on geographical points differs from those of \cite{DrosouP12}, because we consider geographical distance as the basis to calculate diversity between two points.

\begin{dmath}
       \label{eq:harvestine}
       distance(p,p') = acos(cos(p.\mathit{lat}) \times cos(p'.\mathit{lat}) \times cos(p.\mathit{lon})) \times cos(p'.\mathit{lon}) + cos(p.\mathit{lat}) \times sin(p'.\mathit{lat}) \times cos(p.\mathit{lon}) \times sin(p'.\mathit{lon}) + sin(p.\mathit{lat}) \times sin(p'.\mathit{lat})) \times earth\_radius
\end{dmath}

Algorithm \ref{algo:geoh} describes our approach for highlighting $k$ similar and diverse points.
We propose a best-effort greedy approach to efficiently compute highlighted points. We consider an offline step followed by the online execution of our algorithm.

\separateshort
In order to speed up the similarity computation in the online execution, we pre-compute an inverted index for each single point $p \in {\mathcal P}$ in the offline step (as is commonly done in the Web search). Each index ${\mathcal L}_p$ for the point $p$ keeps all other points in ${\mathcal P}$ in decreasing order of their similarity with~$p$.

\separateshort
The first step of Algorithm \ref{algo:geoh} is to find the most similar point to $F$, so-called $p^*$. The point $p^*$ is the closest possible approximation of $F$ in order to exploit pre-computed similarities. The algorithm makes sequential accesses to ${\mathcal L}_{p^*}$ (i.e., the inverted index of the point $p^*$) to greedily maximize diversity. Algorithm \ref{algo:geoh} does not sacrifice efficiency in price of value. We consider a {\em time limit} parameter which determines when the algorithm should stop seeking maximized diversity. Scanning inverted indexes guarantees the similarity maximization even if time limit is chosen to be very restrictive. Our observations with several spatial datasets show that we achieve the diversity of more than $0.9$ with time limit set to $200ms$.


\begin{algorithm}[t]
\DontPrintSemicolon
\KwIn{Points $\mathcal{P}$, Feedback vector $F$, $k$, $\mathit{time\_limit}$}
\KwOut{$\mathcal{P}_k$}
$p^* \gets \mathit{max\_sim\_to}(\mathcal{P},F)$\;
$\mathcal{P}_k \gets \mathit{top\_k}(\mathit{{\mathcal L}_{p^*}},k)$\label{ln:topk}\;
$p_{next} \gets get\_next(\mathit{{\mathcal L}_{p^*}})$\;\label{cd:getnext}
\While{$\mathit{time\_limit}$ $not$ $exceeded$}
       {\label{cd:beginwhile}
       \For{$p_{current} \in {\mathcal P}_k$}
              {
              \If{$\mathit{diversity\_improved}({\mathcal P}_k,p_{next},p_{current})$}
                     {\label{cd:betterdiv}
                     ${\mathcal P}_k \gets \mathit{replace}({\mathcal P}_k,p_{next},p_{current})$\;
                            $break$\;
                     }
              }
              $p_{next} \gets get\_next(\mathit{{\mathcal L}_{p^*}})$\;}\label{cd:endwhile}
       \Return{${\mathcal P}_k$}\; 
       \caption{Get $k$ similar and diverse highlights $\mathit{get\_highlights}()$}
       \label{algo:geoh}
\end{algorithm}

\separateshort
In line \ref{ln:topk} of Algorithm \ref{algo:geoh}, $\mathcal{P}_k$ is initialized with the $k$ highest ranking points in ${\mathcal L}_{p^*}$. Function $get\_next({\mathcal L}_{p^*})$ (line \ref{cd:getnext}) returns the next point $p_{next}$ in ${\mathcal L}_{p^*}$ in sequential order ( as a common practice in information retrieval). Lines \ref{cd:beginwhile} to \ref{cd:endwhile} iterate over the inverted indexes to determine if other points should be considered to increase diversity while staying within the time limit.

\separateshort
The algorithm looks for a candidate point $p_{\mathit current} \in {\mathcal P}_k$ to replace in order to increase diversity. The boolean function $\mathit{diversity\_improved}()$ (line \ref{cd:betterdiv}) checks if by replacing $p_{current}$ by $p_{next}$ in ${\mathcal P}_k$, the overall diversity of the new ${\mathcal P}_k$ increases. It is important to highlight that for each run of the algorithm, we only focus on one specific inverted list associated to the input point. Algorithm \ref{algo:geoh} verifies the similarity  and diversity of each point with all other points, and then processes the normalization.

\section{Experiments}
\label{sec:experiments}

We discuss two sets of experiments. Our first set is on the usefulness of our approach. Then we focus more on discovering IDRs and present few statistics and insights for them. The experiments are done on the a computer with Mac OS operating system, with a 2,8 GHz Intel Core i5. 


\separateshort
First off, we validate the ``usefulness'' of our approach. For this aim, we design a user study with some participants who are all students of Computer Science. Some of them are ``novice'' users who don't know the location under investigation, and some are ``experts.'' Participants should fulfill a task. For each participant, we report a variant of time-to-insight measure, i.e., how long the participants interact with the tool before fulfilling the task. Evidently, less number of interactions are preferred as it means that the participant can reach insights faster.

\separateshort
On the \textsf{Airbnb}\footnote{\it \url{http://insideairbnb.com/get-the-data.html}} dataset of Paris with 1,000 points, we define two different types of tasks: {\em T1: ``finding a point in a requested location''} (e.g., find a home-stay in the ``\textit{Champ de Mars}'' area), and {\em T2: ``finding a point with a requested profile''} (e.g., find a cheap home-stay.) Due to the vagueness associated to these tasks, participants require to go through an exploratory analysis session. Moreover, participants may also begin their navigation either from {\em I1: ``close to the goal''} or {\em I2: ``far from the goal''}.

\begin{table}[t]
\centering
\caption{Interactions of ``novice'' and ``expert'' participants (in seconds)}
\label{tbl:novice}
\begin{tabular}{c|c|c|c|c|}
\cline{2-5}
                                       	& \textbf{T1/I1} 	& \textbf{T2/I1} 	& \textbf{T1/I2}	& \textbf{T2/I2}	\\ \hline
\multicolumn{1}{|c|}{Novices} 				& 1.99            	& 2.38	          	& 2.00              & 2.48              \\ \hline
\multicolumn{1}{|c|}{Experts} 				& 1.72            	& 2.09	          	& 1.70              & 2.14              \\ \hline
\end{tabular}
\end{table}

\separateshort
Table \ref{tbl:novice} shows the results. We observe that on average it takes $2.067$ seconds to achieve defined goals. This shows that implicit feedback capturing is an effective mechanism which helps analysts to reach their goals in a reasonable time. Expert participants need $0.35$ seconds less time on average. Interestingly, starting points, i.e., {\em I1} and {\em I2}, do not have a huge impact on number of steps. It is potentially due to the diversity component which provides distinct options and can quickly guide analyst towards their region of interest. We also observe that the task {\em T2} is an easier task than {\em T1}, as on average it took less to be accomplished. This is potentially due to  where the analyst can request options similar to what she has already observed and greedily move to her preferred regions.

\separateshort
In the second part of our experiments, we employ two different datasets, i.e., \textsf{Airbnb} and \textsf{Yelp}\footnote{\it \url{https://www.yelp.com/dataset}}. We pick a similar subset from both datasets, i.e., home-stays and restaurants in Paris city, respectively. We consider four different sizes of those datasets, i.e., $100$, $1000$, $2000$ and $4000$ points, respectively. For each size of the datasets, we manually perform $20$ sessions, and then we present the results as the average of sessions.

\separateshort
We limit each session to $2$ minutes where we seek for interesting points in the datasets. We capture the following information in each session:

\begin{itemize}[leftmargin=*]
  \item The number of regions created from the mouse moves during the session;
  \item The number of generated IDRs (intersection of regions);
  \item The number of points from the dataset presented in each IDR;
  \item The coverage of points (in the dataset) with IDRs collectively.
\end{itemize}  

\separateshort
Tables \ref{tbl:airbnb} and \ref{tbl:yelp} show the result for \textsf{Airbnb} and \textsf{Yelp}, respectively. In Table \ref{tbl:airbnb}, we observe that the number of regions decreases when the number of points increases. On average, $10$ regions are constructed per session. The average number of points presented in IDRs is $25.97$, which shows that our approach highlights at least $8.05\%$ of points from the dataset, on average.
We notice an outlier in the experiment with 2000 points in Tables \ref{tbl:airbnb}. This happened due the fact that  the analyst concentrated in a very small area generating a smaller number of IDRs, and consequently a smaller number of points.
 
\begin{table*}[t]
\centering
\caption{IDR statistics on \textsf{Airbnb} dataset}
\label{tbl:airbnb}
\begin{tabular}{|c|c|c|c|c|}
\cline{1-5}
\textbf{\# points}  & \textbf{\# regions} 	& \textbf{\# IDRs} 	& \textbf{\# points in IDRs}	& \textbf{\%  points}	\\ \hline
\multicolumn{1}{|c|}{100} 				& 11.35            	& 10.05	          	& 29.40             & 29.40\%            
 \\ \hline
\multicolumn{1}{|c|}{1000} 				& 10.75          	& 6.75	          	& 11.70              & 1.17\%              \\ \hline
\multicolumn{1}{|c|}{2000} 				& 7.37           	& 3.63         	& 5.63             & 0.003\%              \\ \hline
\multicolumn{1}{|c|}{4000} 				& 10.30           	& 10.15	          	& 53.15              & 1.33\%              \\ \hline
\multicolumn{1}{|c|}{\textbf{average}} 				& \textbf{9.94}           	& \textbf{7.64}	          	&\textbf{ 25.97}              & \textbf{8.05\% }             \\ \hline

\end{tabular}
\end{table*}

\separateshort
More uniform results are observed in Table \ref{tbl:yelp}, i.e., for \textsf{Yelp} dataset vis-\`a-vis \textsf{Airbnb}. The average number of generated regions reaches $12.75$ per session. Also, the number of regions decreases by increasing the number of points. The same happens for IDRs, where we obtain an average of $8.9$ IDRs generated per session. The number of points presented in IDRs is on average $108.65$ and it represents on average $13.11\%$ of points highlighted from the dataset.


\begin{table*}[t]
\centering
\caption{IDR statistics on \textsf{Yelp} dataset}
\label{tbl:yelp}
\begin{tabular}{|c|c|c|c|c|}
\cline{1-5}
\textbf{\# points}  & \textbf{\# regions} 	& \textbf{\# IDRs} 	& \textbf{\# points in IDRs}	& \textbf{\%  points}	\\ \hline
\multicolumn{1}{|c|}{100} 				& 14.90            	& 7.55	          	& 28.30             & 28.30\%            
 \\ \hline
\multicolumn{1}{|c|}{1000} 				& 13.90         	& 10.00	          	& 149.55             & 14.96\%              \\ \hline
\multicolumn{1}{|c|}{2000} 				& 11.05         	& 9.80         	& 111.05             & 5.55\%              \\ \hline
\multicolumn{1}{|c|}{4000} 				& 10.45          	& 8.55	          	& 145.7              & 3.64\%              \\ \hline
\multicolumn{1}{|c|}{\textbf{average}} 				& \textbf{12.57}           	& \textbf{8.97}	          	& \textbf{108.65}              & \textbf{13.11\%}              \\ \hline
\end{tabular}
\end{table*}

\section{Related Work}
\label{sec:rel}
To the best of our knowledge, the problem of spatial information highlighting using implicit feedback has not been addressed before in the literature. However, our work relates to few others in their semantics.

\separate
\noindent {\bf Information Highlighting.} The literature contains few instances of information highlighting approaches~\cite{Liang2010,Robinson2011,wongsuphasawat2016voyager,willett2007scented}. However, all these methods are objective, i.e., they 	assume that analyst's preferences are given as a constant input and will never change in the future. This limits their functionality for serving scenarios of exploratory analysis. The only way to fulfill ``spatial guidance'' is to consider the evolutionary and subjective nature of analyst's feedback. In our approach, the feedback vector gets updated in time based on the implicit feedback of the analyst.

\separateshort
Online recommendation approaches can also be considered as an information highlighting approach where recommended items count as highlights. Most recommendation algorithms are space-agnostic and do not take into account the spatial information. While few approaches focus on the spatial dimension~\cite{Bao2015,Levandoski:2012,DrosouP12}, they still lack the evolutionary feedback capturing. Moreover, most recommendation methods miss ``result diversification'', i.e., highlights may not be effective due to overlaps.

\separate
\noindent {\bf Feedback Capturing.} Several approaches are proposed in the state of the art for capturing different forms of feedback~\cite{bhuiyan2012interactive,xin2006discovering,dimitriadou2016aide,kamat2014distributed,Omidvar-Tehrani:2015,boley2013one}. The common approach is a top-$k$ processing methodology in order to prune the search space based on the explicit feedback of the analyst and return a small subset of interesting results of size~$k$. A clear distinction of our proposal is that it doesn't aim for pruning, but leveraging the actual data with potential interesting results (i.e., highlights) that the analyst may miss due to the huge volume of spatial data. Moreover, in a typical top-$k$ processing algorithm, analyst's choices are limited to $k$. On the contrary, our IDR approach enables a freedom of choice where highlights get seamlessly updated with new analyst's choices. 

\separateshort
Few works formulate fusing approaches of explicit and implicit feedbacks to better capture user preferences~\cite{AoidhBW07,Ballatore2008,Liu:2010}. Our approach functions purely on implicit feedback and does not require any sort of explicit signal from the analyst.

\separateshort
\noindent {\bf Region Discovery.} Our approach finds interesting dense regions (IDRs) in order to derive analyst's implicit preferences. There exist several approaches to infer a spatial region for a given set of points \cite{Bevis1989,DUCKHAM2008,FADILI2004,ARAMPATZIS2006,Galton2006,Barber:1996}. The common approach is to cluster points in form of concave and convex polygons. In~\cite{Bevis1989}, an algorithm is proposed to verify if a given point $p$ on the surface of a sphere is located inside, outside, or along the border of an arbitrary spherical polygon. In \cite{DUCKHAM2008,FADILI2004}, a non-convex polygon is constructed from a set of input points on a plane. In \cite{ARAMPATZIS2006,Galton2006}, imprecise regions are delineated into a convex or concave polygon. In our approach, it is important to discover regions by capturing mouse move points. In case a concave polygon is constructed, the ``dents'' of such a polygon may entail points which are not necessarily in $\mathcal{M}$. In the IDR's algorithm, however, we adapt Quickhull~\cite{Barber:1996}, due its simplicity, efficiency and its natural implementation of convex polygons.

\section{Limitations}
\label{sec:limitations}

 In this paper, we presented a solution for highlighting out-of-sight information using a polygon-based approach for capturing implicit feedbacks. To the best of our knowledge, our work is the first effort towards formalizing and implementing information highlighting using implicit feedback. However, we consider our work as an on-going effort where we envision to address some limitations in the future, such as ``customizability'', ``performance'', ``cold start'', and ``quantitative experiments''.
 
 \separateshort
In this section we present some limitations of our proposed work, describing what we will consider as future work. One limitation is about the ``customizable'' use of geographical maps as an interaction means. As we only consider static maps, we plan to work on translations and rotations as a future work. Another gap that we envision to work on is  performance. We plan to run an extensive performance study to detect bottlenecks of our approach.

\separateshort
Our problem bears similarities with recommendation algorithms where the quality of the output may be influenced by scarce availability of input. This problem is referred to as the cold start problem~\cite{LeroyCB10}. While there is no guarantee for a meaningful highlight in case of the complete absence of implicit feedbacks, our approach can return a reasonable set of highlights even with one single iteration of mouse moves. In the future, we envision to tackle the no-input challenge by leveraging statistical properties of the spatial data to obtain a default view for highlights.

\separateshort
Another limitation is the medium-size datasets to be processed. Our algorithm processes similarity and diversity in an $O(n^2)$ complexity. Also Quickhull~\cite{Barber:1996} uses a divide and conquer approach similar to that of Quicksort, and its worst complexity is $O(n^2)$. While processing a 10K-point dataset is straightforward in our framework, we plan to experiment with larger datasets in the future by improving our algorithms towards better performance. Another direction for future work is to consider experiments which measure the quantitative and qualitative influence of each component separately.  

\section{Conclusion and Future Work}
\label{sec:conclusion}

In this paper, we present an approach to explore Interesting Dense Regions (IDRs) using implicit feedback in order to detect analyst latent preferences. The implicit feedbacks are captured from mouse moves of analysts over the geographical map while analyzing spatial data. We formalize a novel polygon-based mining algorithm which returns few highlights in-line with analyst's implicit preferences. The highlights enable analysts to focus on what matters the most and prevent information overload.

\separateshort
We consider various future directions for this work. First, we are interested to incorporate an ``explainability'' component which can describe causalities behind preferences. For instance, we are interested to find seasonal patterns to see why the preferences of analysts change from place to place during various seasons of the year. Another direction is to incorporate ``Query by Visualization'' approaches, where analysts can specify their intents alongside their implicit preferences, directly on the map~\cite{siddiqui2016effortless}.



\end{document}